\documentclass[aps,prb,showpacs,twocolumn]{revtex4}
\usepackage{graphicx}
\begin{document}
\title{Radiation-induced magnetoresistance oscillation
in a two-dimensional electron gas\\ in Faraday geometry}
\author{X.L. Lei and S.Y. Liu}
\affiliation{Department of Physics, Shanghai Jiaotong University,
1954 Huashan Road, Shanghai 200030, China}

\begin{abstract}
Microwave-radiation induced giant magnetoresistance oscillations recently discovered
in high-mobility two-dimensional electron systems
are analyzed theoretically. Multiphoton-assisted impurity scatterings
are shown to be the primary origin of the oscillation.
Based on a theory which considers the interaction of electrons with electromagnetic
fields and the effect of the cyclotron resonance in Faraday geometry, 
we are able not only to reproduce the correct period, phase and the negative 
resistivity of the main oscillation, but also to predict the secondary peaks 
and additional maxima and minima observed in the experiments. These peak-valley
structures are identified to relate respectively to single-, double- and 
triple-photon processes. 

\end{abstract}


\maketitle

The discovery of a new type of giant magnetorersistance oscillations 
in a high mobility two-dimensional (2D) electron gas (EG) subject to 
crossed microwave (MW) radiation field and a magnetic 
field,\cite{Zud01,Ye,Mani,Zud03,Mani05507}
especially the observation of "zero-resistance" states 
developed from the oscillation minima,\cite{Mani,Zud03,Yang,Dor}
has revived tremendous interest in magneto-transport in 2D electron systems.\cite{
Durst,Andreev,Anderson,Xie,Phil,Koul} These radiation-induced oscillations 
of longitudinal resistivity $R_{xx}$ are accurately periodical in inverse 
magnetic field $1/B$ with period determined  
by the MW frequency $\omega$ rather than the electron density $N_{\rm e}$.
The observed $R_{xx}$ oscillations exhibit a smooth 
magnetic-field variation with resistivity maxima at $\omega/\omega_c=j-\delta_{-}$ 
and minima at $\omega/\omega_c=j+\delta_{+}$ ($\omega_c$ is the cyclotron
frequency, $j=1,2,3...$) having positive $\delta_{\pm}$ 
around $\frac{1}{4}$.\cite{Mani} 
The resistivity minimum goes downward 
with increasing sample mobility and/or increasing radiation intensity 
until a "zero-resistance" state shows up, while the Hall resistivity 
keeps the classical form $R_{xy}=B/N_{\rm e}e$ 
with no sign of quantum Hall plateau 
over the whole magnetic field range exhibiting $R_{xx}$ oscillation.

To explore the origin of the peculiar "zero-resistance" states, 
different mechanisms have been suggested.\cite{Durst,Andreev,Anderson,Xie,Phil,Koul}  
It is understood that the appearance of negative longitudinal resistivity 
or conductivity in a uniform model suffices to explain the observed vanishing 
resistance.\cite{Andreev} 
The possibility of absolute negative photoconductance in a 2DEG subject to a 
perpendicular
magnetic field was first explored 30 years ago by Ryzhii.\cite{Ryz,Ryz86} 
Recent works\cite{Durst,Anderson,Xie} indicated that the periodical 
structure of the density of states (DOS) of the 2DEG in a magnetic field 
and the photon-excited electron scatterings are the origin of 
the magnetoresistance oscillations. 
Durst {\it et al.}\cite{Durst} presented a microscopic analysis 
for the conductivity assuming a $\delta$-correlated disorder and a simple  
form of the 2D electron self-energy oscillating with the magnetic field, 
obtaining the correct period, phase and the possible negative resistivity. 
Shi and Xie\cite{Xie} reported a similar result using the Tien and Gorden
current formula\cite{Tien} for photon-assisted coherent tunneling. 
In these studies, however,
the magnetic field is to provide an oscillatory DOS only and 
the high frequency (HF) field enters as if there is no magnetic field 
or with a magnetic field in Voigt configuration. The experimental setup requires  
to deal with the magnetic field ${\bf B}$ perpendicular to the HF electric field. 
In this Faraday configuration, 
the electron moving due to HF field, experiences a Lorentz force
which gives rise to additional electron motion in the perpendicular direction. 
In the range of $\omega\sim\omega_c$, the electron velocities in both directions are 
of the same order of magnitude and are resonantly enhanced. 
This cyclotron resonance (CR) of the HF current response will certainly
change the way how the photons assist 
the electron scattering.
   
In this Letter, we construct a microscopic model for the interaction of electrons 
with electromagnetic fields in Faraday geometry. The basic idea is that, 
under the influence of a spatially uniform HF electric field, 
the center-of-mass (CM) of the whole 2DEG in a magnetic field performs a cyclotron 
motion modulated by the HF field of frequency $\omega$. 
In an electron gas having impurity and/or phonon scatterings, there exist 
couplings between this CM motion and the relative motion of the 2D electrons. 
It is through these couplings that a spatially uniform HF electric field affects 
the relative motion of electrons by opening additional channels for 
electron transition between different states. 
Based on the theory for photon-assisted magnetotransport developed from this 
physical idea, we show that the main experimental results of the radiation-induced 
magnetoresistance oscillations can be well reproduced.
We also obtain the secondary peaks and additional maxima and minima observed
in the experiments.\cite{Zud03,Dor}  

For a general treatment,
we consider $N_{\rm e}$ electrons in a unit area of a quasi-2D system
in the $x$-$y$ plane with a confining potential $V(z)$ in the $z$-direction. 
These electrons, besides interacting with each other, are scattered by
random impurities/disorders and by phonons in the lattice. 
To include possible elliptically polarized MW illumination we assume that 
a uniform dc electric field ${\bf E}_0$ and ac field 
${\bf E}_t\equiv{\bf E}_s \sin(\omega t)+{\bf E}_c\cos(\omega t)$ 
of frequency $\omega$ are applied in the $x$-$y$ plane, 
together with a magnetic field ${\bf B}=(0,0,B)$ along the $z$ direction.
In terms of the 2D CM momentum and coordinate of
the electron system,\cite{Lei85,Lei851,Ting} which are defined as 
${\bf P}\equiv\sum_j {\bf p}_{j\|}$ 
and ${\bf R}\equiv N_{\rm e}^{-1}\sum_j {\bf r}_{j\|}$  
with ${\bf p}_{j\|}\equiv(p_{jx},p_{jy})$ and ${\bf r}_{j\|}\equiv (x_j,y_j)$
being the momentum and coordinate of the $j$th electron in the 2D plane,
and the relative electron momentum and coordinate 
${\bf p}_{j\|}'\equiv{\bf p}_{j\|}-{\bf P}/N_{\rm e}$ and 
${\bf r}_{j\|}'\equiv{\bf r}_{j\|}-{\bf R}$,
the Hamiltonian of the system can be written as the sum of 
a CM part $H_{\rm cm}$
and a relative-motion part $H_{\rm er}$ 
(${\bf A}({\bf r})$ is the vector potential of the ${\bf B}$ field),
\begin{eqnarray}
H_{\rm cm}=\frac 1{2N_{\rm e}m}({\bf P}-N_{\rm e}e{\bf A}({\bf
R}))^2-N_{\rm e}e({\bf E}_{0}+{\bf E}_t)\cdot {\bf R},&&\\
H_{\rm er}=\sum_{j}\Big[\frac{1}{2m}\left({\bf p}_{j\|}'-e{\bf A}
({\bf r}_{j\|}')\right)^{2}
+\frac{p_{jz}^2}{2m_z}+V(z_j)\Big]\,\,&&\nonumber\\
+\sum_{i<j}V_c({\bf r}_{i\|}'-{\bf r}_{j\|}',z_i,z_j),\,\,\,\,\,\,&&
\end{eqnarray}
together with couplings of electrons to impurities and phonons,
$H_{\rm ei}$ and $H_{\rm ep}$. Here $m$ and $m_z$ are respectively the electron 
effective mass parallel and perpendicular to the plane, and $V_c$ stands for the 
electron-electron Coulomb interaction. 
Note that although in $H_{\rm cm}$ and $H_{\rm er}$ the CM and the relative 
electron motion are completely separated, the CM coordinate ${\bf R}$
enters $H_{\rm ei}$ and $H_{\rm ep}$.\cite{Lei85,Lei851} Starting from
the Heisenberg operator equations for the rate of change of the CM velocity 
$\dot{\bf V}=-i[{\bf V},H]+\partial{\bf V}/\partial t$ with ${\bf V}=-i[{\bf R},H]$,
and that of the relative electron energy 
$\dot{H}_{\rm er}=-{\rm i}[H_{\rm er},H]$, we proceed with the determination of their
statistical averages.

The CM coordinate ${\bf R}$ 
and velocity ${\bf V}$ in these equations can be treated classically, i.e. 
as the time-dependent
expectation values of the CM coordinate and velocity,\cite{Lei85}
${\bf R}(t)$ and ${\bf V}(t)$, such that ${\bf R}(t)-{\bf R}(t^{\prime})
=\int_{t^{\prime}}^t{\bf V}(s)ds$.
We are concerned with the steady transport
under an irradiation of single frequency and focus on the 
photon-induced dc resistivity and the energy absorption of the HF field. 
These quantities are directly related to the time-averaged and/or base-frequency 
oscillating components of the CM velocity.
At the same time, in an ordinary semiconductor the effect of higher harmonic current 
is safely negligible for the HF field intensity in the experiments. 
Hence, it suffices to assume that the CM 
velocity, i.e. the electron drift velocity, consists of a dc
part ${\bf v}_0$ and a stationary time-dependent part that
\begin{equation}
{\bf V}(t)={\bf v}_0+{\bf v}_1 \cos(\omega t)+{\bf v}_2 \sin(\omega t).
\end{equation}
This time-dependent CM velocity enters all the operator equations having 
couplings to impurities and/or phonons 
in the form of the following exponential factor, which can be expanded 
in terms of Bessel functions ${\rm J}_n(x)$:
\begin{eqnarray}
\hspace*{-1.3cm}&&{\rm e}^{-{\rm i}{\bf q}\cdot \int_{t^{\prime }}^{t}{\bf V}(s)ds}
=\sum_{n=-\infty }^{\infty }{\rm J}_{n}^{2}(\xi ){\rm e}^{{\rm i}({\bf q}\cdot {\bf 
v}_0-n\omega) (t-t^{\prime
})}+\nonumber\\
&&\sum_{m\neq 0}{\rm e}^{{\rm i}m(\omega t-\varphi )}\sum_{n=-\infty }^{\infty
}{\rm J}_{n}(\xi ){\rm J}_{n-m}(\xi ){\rm e}^{{\rm i}({\bf q}\cdot {\bf v}_0-n\omega) 
(t-t^{\prime })}.\nonumber
\end{eqnarray}
Here $\xi\equiv \sqrt{({\bf q}_\|\cdot {\bf v}_1)^2+({\bf q}_\|\cdot {\bf 
v}_2)^2}/\omega$
and  
$\tan \varphi=({\bf q}\cdot {\bf v}_2)/({\bf q}\cdot {\bf v}_1)$.
 On the other hand, for 2D systems having 
electron sheet density of order of 10$^{15}$ m$^{-2}$, 
the intra-band and inter-band Coulomb interactions are sufficiently strong 
that it is adequate to describe the relative-electron transport state using a single 
electron temperature $T_{\rm e}$. Except this, the electron-electron interaction
is treated only in a mean-field level under random phase approximation 
(RPA).\cite{Lei85,Lei851} 
For the determination of unknown parameters ${\bf v}_0$, ${\bf v}_1$, ${\bf v}_2$, 
and ${T_{\rm e}}$, it suffices to know the damping force 
up to the base frequency oscillating term 
${\bf F}(t)= {\bf F}_0+{\bf F}_s\sin(\omega t)+{\bf F}_c\cos(\omega t)$, 
and the energy-related 
quantities up to the time-average term. We finally obtain the following 
force and energy balance equations:
\begin{eqnarray}
0&=&N_{\rm e}e{\bf E}_{0}+N_{\rm e} e ({\bf v}_0 \times {\bf B})+
{\bf F}_0,\label{eqv0}\\
{\bf v}_{1}&=&\frac{e{\bf E}_s}{m\omega}+\frac{{\bf F}_s}{N_{\rm e}m\omega }
-\frac{e}{m\omega }({\bf v}_{2}\times
{\bf B}),\label{eqv1}\\
-{\bf v}_{2}&=&\frac{e{\bf E}_c}{m\omega}+\frac{{\bf F}_c}{N_{\rm e}m\omega }
-\frac{e}{m\omega }({\bf v}_{1}
\times {\bf B}),\label{eqv2}
\end{eqnarray}
\begin{equation}
N_{\rm e}e{\bf E}_0\cdot {\bf v}_0+S_{\rm p}- W=0.
\label{eqsw}
\end{equation}
Here
\begin{eqnarray}
{\bf F}_{0}=\sum_{{\bf q}_\|}\left| U({\bf q}_\|%
)\right| ^{2}%
\sum_{n=-\infty }^{\infty }{\bf q}_\|{\rm J}_{n}^{2}(\xi )\Pi _{2}({\bf %
q}_\|,\omega_0-n\omega )\,\,\,\,\,&&\nonumber\\
+\sum_{{\bf q}}\left| M({\bf q})\right|
^{2}\sum_{n=-\infty
}^{\infty }{\bf q}_\|{\rm J}_{n}^{2}(\xi )\Lambda _{2}({\bf q},\omega_0+\Omega _{{\bf 
q}}-n\omega )&&
 \label{eqf0}
\end{eqnarray}
is the time-averaged damping force, $S_{\rm p}$ is the time-averaged rate of the 
electron energy-gain from the HF field,
$\frac{1}{2}N_{\rm e}e({\bf E}_s\cdot{\bf v}_2+{\bf E}_c\cdot{\bf v}_1)$,
which can be written in a form obtained 
from the right hand side of Eq.\,(\ref{eqf0}) by replacing the ${\bf q}_\|$ factor with 
$n \omega$, 
and $W$ is the time-averaged rate of the electron energy-loss due to coupling 
with phonons, whose expression can be obtained from the second term on the right hand 
side of Eq.\,(\ref{eqf0}) by replacing the ${\bf q}_\|$ factor 
with $\Omega_{\bf q}$, the energy of a wavevector-${\bf q}$ phonon. 
The oscillating frictional force amplitudes 
${\bf F}_s\equiv {\bf F}_{22}-{\bf F}_{11}$ and 
${\bf F}_c\equiv {\bf F}_{21}+{\bf F}_{12}$ are given by ($\mu=1,2$)
\begin{eqnarray}
{\bf F}_{1\mu}=-\sum _{{\bf q}_\|}{\bf q}_\|\eta_{\mu}| U({\bf q}_\|%
)| ^{2}\sum_{n=-%
\infty }^{\infty }\left[ {\rm J}_{n}^{2}(\xi )\right] ^{\prime }\Pi _{1}(%
{\bf q}_\|,\omega_0-n\omega )&&\nonumber\\
- 
\sum_{\bf q}{\bf q}_\|\eta_{\mu}| M({\bf q})|
^{2}\sum_{n=-\infty
}^{\infty }\left[ {\rm J}_{n}^{2}(\xi )\right] ^{\prime }\Lambda _{1}({\bf q%
}, \omega_0+\Omega _{{\bf q}}-n\omega ),&&\nonumber\\ 
{\bf F}_{2\mu}=\sum_{{\bf q}_\|}{\bf q}_\|\frac{\eta_{\mu}}
{\xi}| U({\bf q}_\|)| ^{2}%
\sum_{n=-\infty }^{\infty }2n{\rm J}_{n}^{2}(\xi )\Pi _{2}({\bf %
q}_\|,\omega_0-n\omega )\,\,&&\nonumber\\
+ 
\sum_{{\bf q}}{\bf q}_\|\frac{\eta_{\mu}}{\xi}| M({\bf q})|^{2}\sum_{n=-\infty
}^{\infty }2n{\rm J}_{n}^{2}(\xi )\Lambda _{2}({\bf q},\omega_0+\Omega _{\bf q}-n\omega 
).
&&\nonumber
 \end{eqnarray}
In these expressions,
$\eta_{\mu}\equiv {\bf q}_\|\cdot {\bf v}_{\mu}/\omega \xi$;
$\omega_0\equiv {\bf q}_\|\cdot {\bf v}_0$;
$U({\bf q}_\|)$ and $M({\bf q})$ stand for effective impurity and phonon
scattering potentials, 
$\Pi_2({\bf q}_\|,\Omega)$ and
$
\Lambda_2({\bf q},\Omega)=2\Pi_2({\bf q}_\|,\Omega)
[n(\Omega_{\bf q}/T)-n(\Omega/T_{\rm e})]
$\,(with $n(x)\equiv 1/({\rm e}^x-1)$)
are the imaginary parts of the electron density correlation function 
and electron-phonon correlation function in the presence of the magnetic field.
$\Pi_1({\bf q}_\|,\Omega)$ and  $\Lambda_1({\bf q},\Omega)$
are the real parts of these two correlation functions.
The effect of interparticle Coulomb interactions are included in them 
to the degree of level broadening and RPA screening.

The HF field enters through the argument $\xi$ of the Bessel functions 
in ${\bf F}_0$, ${\bf F}_{\mu\nu}$, $W$ and 
$S_{\rm p}$. Compared with that without the HF field 
($n=0$ term only),\cite{Lei98} we see that in an electron gas 
having impurity and/or phonon scattering (otherwise homogeneous),
a HF field of frequency $\omega$ opens additional channels for electron 
transition: an electron in a state can absorb or emit one or several photons
and scattered to a different state with the help of impurities and/or phonons.
The sum over $|n|\geq 1$ represents contributions of single and multiple
photon processes of frequency-$\omega$ photons. 
These photon-assisted scatterings help to transfer
energy from the HF field to the electron system ($S_{\rm p}$) 
and give rise to additional damping force on the moving electrons. 
Note that ${\bf v_1}$ and ${\bf v}_2$ 
always exhibit CR in the range $\omega\sim\omega_c$, 
as can be seen from Eqs.(\ref{eqv1}) and (\ref{eqv2}) rewritten in the form
\begin{eqnarray}
&&{\bf v}_{1}=(1-{\omega_c^2}/{\omega^2})^{-1}\left\{
\frac{e}{m\omega}\left[{\bf E}_s+\frac{e}{m\omega }(
{\bf E}_c\times{\bf B})\right]\right.\nonumber\\
&&\hspace{1.0cm}+\left.\frac{1}{N_{\rm e}m\omega }
\left[{\bf F}_s+
\frac{e}{m\omega }(
{\bf F}_c\times
{\bf B})\right]\right\},\label{vv1}\\
&&{\bf v}_{2}=({\omega_c^2}/{\omega^2}-1)^{-1}\left \{
\frac{e}{m\omega}\left[{\bf E}_c-\frac{e}{m\omega }(
{\bf E}_s\times{\bf B})\right]\right.\nonumber\\
&&\hspace{1.0cm}+\left.\frac{1}{N_{\rm e}m\omega }
\left[{\bf F}_c-
\frac{e}{m\omega }
({\bf F}_s\times
{\bf B})\right]\right\}.\label{vv2}
\end{eqnarray}
Therefore, $\xi$ may be significantly different from the argument of the 
corresponding Bessel functions in the case without a magnetic field or 
with a magnetic field in Voigt configuration.\cite{Lei98} 

Eqs.\,(\ref{eqv0})-(\ref{eqsw}) can be used to describe the transport 
and optical properties of magnetically-biased quasi-2D semiconductors 
subject to a dc field and a HF field. 
Taking ${\bf v}_0=(v_{0x},0,0)$ in the $x$ direction,
 Eq.\,(\ref{eqv0}) yields
transverse resistivity
$R_{xy}\equiv E_{0y}/N_{\rm e}ev_{0x}=B/N_{\rm e}e$, and 
longitudinal resistivity $
R_{xx}\equiv E_{0x}/N_{\rm e}ev_{0x}=-F_0/N_{\rm e}^2e^2v_{0x}$.
The linear magnetoresistivity is then 
\begin{eqnarray}
R_{xx}&=&-\sum_{{\bf q}_\|}q_x^2\frac{|
U({\bf q}_\|)| ^2}{N_{\rm e}^2 e^2}\sum_{n=-\infty }^\infty {\rm J}_n^2(\xi)\left. 
\frac {\partial \Pi_2}{\partial\, \Omega }\right|_{\Omega =n\omega }\nonumber\\
&&- \sum_{ {\bf q}} q_x^2\frac{\left| M ( {\bf
q})\right| ^2}{N_{\rm e}^2 e^2}\sum_{n=-\infty }^\infty {\rm J}_n^2(\xi)\left. 
\frac {\partial \Lambda_2}{\partial\, \Omega }\right|_{\Omega =\Omega_{{\bf 
q}}+n\omega}.
\,\,\,\,\,\,\,\label{rxx}
\end{eqnarray}
The parameters ${\bf v}_1$, ${\bf v}_2$ and $T_{\rm e}$ in (\ref{rxx}) 
should be determined by solving equations (\ref{eqv1}), (\ref{eqv2})
and (\ref{eqsw}) with vanishing ${\bf v}_0$. We see that although 
the transverse resistivity $R_{xy}$ remains the classical form, 
the longitudinal resistivity $R_{xx}$ can be strongly affected by the irradiation. 

We calculate the unscreened $\Pi_2({\bf q}_{\|}, \Omega)$ function of the 2D
system in a magnetic field by means of Landau representation:\cite{Ting}
\begin{eqnarray}
&&\hspace{-0.7cm}\Pi _2({\bf q}_{\|},\Omega ) =  \frac 1{2\pi
l_{\rm B}^2}\sum_{n,n'}C_{n,n'}(l_{\rm B}^2q_{\|}^2/2) 
\Pi _2(n,n',\Omega),
\label{pi_2}\\
&&\hspace{-0.7cm}\Pi _2(n,n',\Omega)=-\frac2\pi \int d\varepsilon
\left [ f(\varepsilon )- f(\varepsilon +\Omega)\right ]\nonumber\\
&&\,\hspace{2cm}\times\,\,{\rm Im}G_n(\varepsilon +\Omega){\rm Im}G_{n'}(\varepsilon ),
\end{eqnarray}
where $l_{\rm B}=\sqrt{1/|eB|}$ is the magnetic length,
$
C_{n,n+l}(Y)\equiv n![(n+l)!]^{-1}Y^le^{-Y}[L_n^l(Y)]^2
$
with $L_n^l(Y)$ the associate Laguerre polynomial, $f(\varepsilon
)=\{\exp [(\varepsilon -\mu)/T_{\rm e}]+1\}^{-1}$ the Fermi distribution
function, and ${\rm Im}G_n(\varepsilon )$ is the imaginary part of the
Green's function, or the DOS, of the Landau level $n$.
The real part functions $\Pi_1({\bf q}_{\|},\Omega)$ and 
$\Lambda_1({\bf q}_{\|},\Omega)$ can be obtained from their  
imaginary parts via the Kramers-Kronig relation.

In principle, to obtain the Green's function $G_n(\varepsilon )$,
a self-consistent calculation has to be carried out with all the scattering mechanisms 
included.\cite{Leadley}
In this Letter we do not attempt a self-consistent calculation of
$G_n(\varepsilon)$ but choose a Gaussian-type function
for the purpose of demonstrating the $R_{xx}$ oscillations ($\varepsilon_n$
is the energy of the $n$-th Landau level):\cite{Ando82}
\begin{equation}
{\rm Im}G_n(\varepsilon)=-(\pi/2)^{\frac{1}{2}}\Gamma^{-1}
\exp[-(\varepsilon-\varepsilon_n)^2/(2\Gamma^2)]
\end{equation}
with a broadening parameter $\Gamma=(2e\omega_c\alpha/\pi m \mu_0)^{1/2}$,
where $\mu_0$ is the linear mobility at temperature $T$ 
in the absence of the magnetic field 
and $\alpha > 1$ is a semiempirical parameter to take account the difference 
of the transport scattering time determining the mobility $\mu_0$, 
from the single particle lifetime.\cite{Mani,Durst,Anderson}
\begin{figure}
\includegraphics [width=0.45\textwidth,clip=on] {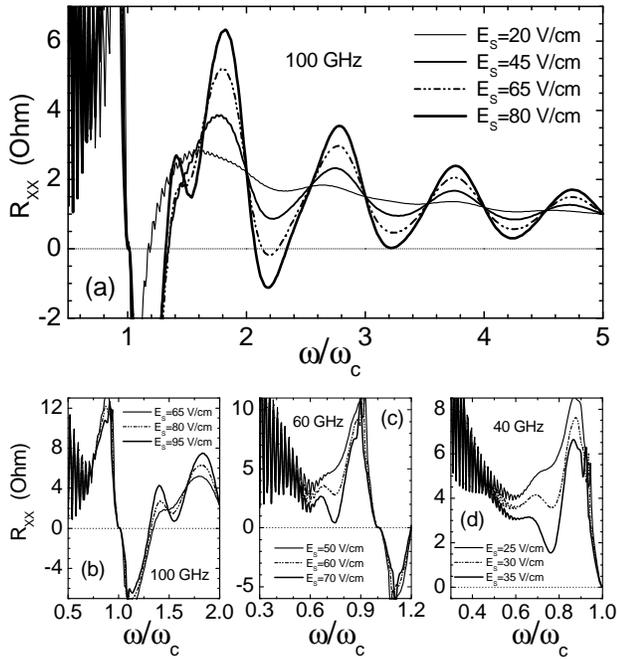}
\vspace*{-0.2cm}
\caption{The longitudinal magnetoresistivity $R_{xx}$ of a GaAs-based 2DEG 
subject to a lineraly polarized HF field $E_s\sin(\omega t)$. The parameters are: 
temperature 
$T=1$\,K, electron density $N_{\rm e}=3.0\times 10^{11}$\,cm$^{-2}$, 
zero-magnetic-field linear mobility $\mu_0=2.4\times 10^7$\,cm$^2$ V$^{-1}$ s$^{-1}$, 
and the broadening coefficient $\alpha=12$.}
\label{fig1}
\end{figure}

The moderate microwave intensity for the $R_{xx}$ oscillation in
these high-mobility samples
yield only slight electron heating, which is unimportant as far as the 
main phenomenon is concerned and is neglected for simplicity.
We consider scatterings from remote impurities 
as well as from acoustic phonons. 
After solving ${\bf v}_1$ and ${\bf v}_2$ from Eqs.\,(\ref{vv1}) and(\ref{vv2}) the 
magnetoresisivity $R_{xx}$ can   
be obtained directly from Eq.\,(\ref{rxx}). At lattice temperature $T=1$\,K, the 
contribution from photon-assisted phonon scattering is minor. The role of acoustic 
phonons, however, becomes essential at elevated lattice temperatures.
Calculations were carried out for linearly polarized MW fields  
with multiphoton processes included.
  
Fig.\,1 shows the calculated longitudinal resistivity $R_{xx}$ 
versus $\omega/\omega_c\equiv \gamma_c$ subject to a linearly polarized 
MW radiation of frequency 
$\omega/2\pi=100$\,GHz at four values of amplitude: 
$E_s=20, 45, 65$ and 80\,V/cm. Shubnikov-de Haas (SdH)
oscillations show up strongly at high $\omega_c$ side, and gradually 
decay away as $1/\omega_c$ increases.
All resistivity curves exhibit pronounced oscillation having main oscillation period 
$\gamma_c=1$ (they are crossing at integer points $\gamma_c=2,3,4,5$).
The resistivity maxima locate around $\gamma_c=j-\delta_{-}$ 
and minima around $\gamma_c=j+\delta_{+}$ with $\delta_{\pm}\sim 0.23-0.25$
for $j=3,4,5$, $\delta_{\pm}\sim 0.17-0.21$ for $j=2$, 
and even smaller $\delta_{\pm}$ for $j=1$.
The magnitude of the oscillation increases with increasing
HF field intensity for $\gamma_c >1.5$. Resistivity gets into negative value for 
$E_s=80$\,V/cm around the minima at $j=1,2$ and 3, for $E_s=65$\,V/cm at $j=1$ and 2,
and for $E_s=20$ and 45\,V/cm at $j=1$.  
All these features, which are in fairly good agreement with experimental
findings,\cite{Zud01,Mani05507,Mani,Zud03} are relevant mainly to single-photon 
($|n|=1$) processes.
Anomaly appears in the vicinity 
of $\gamma_c=1$, where the CR greatly enhances the effective amplitude 
of the HF field in photon-assisted scatterings and multiphoton processes show up.
The amplitudes of the $j=1$ maximum and minimum no longer monotonically change with
field intensity. Furthermore, there appears a shoulder around 
$\gamma_c=1.5$ on the curves of $E_s=45$ and 65\,V/cm, and it develops into
a secondary peak in the $E_s=80$\,V/cm case. This has already been seen 
in the experiment (Fig.\,2 in Ref.\,\onlinecite{Zud03}). The valley between
 $\gamma_c=1.4$ and 1.8 peaks can descend down to negative as $E_s$ 
 increasing further (Fig.\,1b). The appearance of the secondary peak is 
 due to two-photon ($|n|= 2$) processes.

Radiation-induced resistivity behavior at $\gamma_c<1$ is shown more clearly
in the $\omega/2\pi=60$\,GHz case. As seen in Fig.\,1c, a shoulder 
around $\gamma_c=$\,0.4-0.6 with a minmum at $\gamma_c=0.6$ can be 
indentified from the SdH oscillation background for all three curves, which is
related to two-photon process. With increasing MW strength 
there appears a clear peak around $\gamma_c=0.68$ and a valley around $\gamma_c=0.76$. 
This peak-valley is mainly due to
three-photon ($|n|=3$) process. In the case of $40$\,GHz,
similar peak and valley also show up (Fig.\,1d). 

Qualitatively, the main oscillation features 
come from the symmetrical property 
of the DOS function in a magnetic field. Since 
$G_n(\varepsilon-j\omega_c)=G_{n-j}(\varepsilon)$ for any integer $j$, the impurity 
contribution to $R_{xx}$ from the $n$-photon process, 
which is related to the weighted summation 
of the derivative $\Pi_2$ function over all the Landau levels at frequency $n\omega$
 [Eq.\,(\ref{rxx})], has an intrinsic periodicity characterized by $n\omega=j\omega_c$.  
The main oscillation of $R_{xx}$ shown in Fig.\,1a relates to single-photon process and 
characterized by $\omega=j\omega_c$. 
We have also performed calculation using a Lorentz-type DOS function
and find that, although the oscillating amplitude and the exact peak and valley
positions are somewhat different, the basic periodic feature of the radiation-induced 
magnetoresistivity oscillation remains.

This work was supported by the National Science Foundation of China,
the Special Funds for Major State Basic Research Project, and
the Shanghai Municipal Commission of Science and Technology.

\end{document}